# The Lorentz and Levi-Civita Conservation Laws Prohibit the Existence of Black Holes

**Fang-Pei Chen**

Department of Physics, Dalian University of Technology, Dalian 116024, China.

E-mail: chenfap@dlut.edu.cn

**Abstract** To propose that black holes do not exist would be the fundamental way to resolve the contradiction between event horizons and quantum mechanics. In this paper we shall use the Lorentz and Levi-Civita conservation laws to explain how an event horizon might not exist, and the reasons for its nonexistence are presented by using the theory of classical gravity.

## 1. Introduction

Recently, based on comprehensive studies on the quantum properties of black holes, it has been suggested by some physicists that black holes do not exist [1, 2]. Chapline has pointed out [1] clearly that "event horizons cannot exist in the real world for the simple reason that they are inconsistent with quantum mechanics". On one hand "an event horizon makes it impossible to everywhere synchronize atomic clocks", and on the other hand "quantum mechanics requires a universal time". Krauss, Vachaspti and others [2] have conducted a series of mathematical calculations which concluded that black holes cannot exist. In my opinion, their conclusions need to be supported by a broader range of theoretical considerations than the quantum mechanics alone. It is particularly important to find an explanation from the theory of classical gravity.

In this paper the Lorentz and Levi-Civita's conservation laws [3] shall be used to explain how an event horizon might not exist and the explanations for its nonexistence are based on a rigorous theoretical analysis by using the theory of classical gravity.

## 2. The Lorentz and Levi-Civita's conservation laws

For many theories of gravitation, there are two conservation laws of energy-momentum tensor density [3, 4]:

$$\frac{\partial}{\partial x^\lambda}(\sqrt{-g}\,T^\lambda_{(M)\alpha} + \sqrt{-g}\,T^\lambda_{(G)\alpha}) = 0 \tag{1}$$

$$\sqrt{-g}\,T^{\lambda}_{(M)\alpha} + \sqrt{-g}\,T^{\lambda}_{(G)\alpha} = 0 \qquad (2)$$

and
$$\frac{\partial}{\partial x^{\lambda}}\left(\sqrt{-g}\,T^{\lambda}_{(M)\alpha} + \sqrt{-g}\,\tilde{t}^{\lambda}_{(G)\alpha}\right) = 0 \qquad (3)$$

$$\sqrt{-g}\,\tilde{t}^{\lambda}_{(G)\alpha} = \sqrt{-g}\,T^{\lambda}_{(G)\alpha} - \frac{\partial}{\partial x^{\beta}} u^{\lambda\beta}_{(G)\alpha} \quad \left(\frac{\partial}{\partial x^{\beta}} u^{\lambda\beta}_{(G)\alpha} = -\frac{\partial}{\partial x^{\beta}} u^{\beta\lambda}_{(G)\alpha}\right) \qquad (4)$$

where
$$\sqrt{-g}\,T^{\lambda}_{(M)\alpha} \stackrel{def}{=} h^{i}_{.\alpha}\,\frac{\delta(\sqrt{-g}\,L_M)}{\delta h^{i}_{.\lambda}}, \qquad \sqrt{-g}\,T^{\lambda}_{(G)\alpha} \stackrel{def}{=} h^{i}_{.\alpha}\,\frac{\delta(\sqrt{-g}\,L_G)}{\delta h^{i}_{.\lambda}}.$$

Eqs. (1, 2) are called Lorentz and Levi-Civita's conservation laws and Eqs. (3, 4) are called Einstein's conservation laws [3, 5]. These two conservation laws have the same energy-momentum tensor density $\sqrt{-g}\,T^{\lambda}_{(M)\alpha}$ for matter field, but have different energy-momentum tensor density or energy-momentum pseudo tensor density, i.e. $\sqrt{-g}\,T^{\lambda}_{(G)\alpha}$ or $\sqrt{-g}\,\tilde{t}^{\lambda}_{(G)\alpha}$ for gravitational field. $\sqrt{-g}\,T^{\lambda}_{(M)\alpha}$ and $\sqrt{-g}\,T^{\lambda}_{(G)\alpha}$ are tensor densities but $\sqrt{-g}\,\tilde{t}^{\lambda}_{(G)\alpha}$ is pseudo tensor density, so that Eqs. (1, 2) are covariant relations but Eqs. (3, 4) are not. $\sqrt{-g}\,T^{\lambda}_{(G)\alpha}$ and $\sqrt{-g}\,\tilde{t}^{\lambda}_{(G)\alpha}$ are linked by Eq. (4), so they have the equivalence relation [6] in mathematical sense.

In the last few years the author has thoroughly studied Lorentz and Levi-Civita's conservation laws and found that these conservation laws have a broad range of applications [3, 6-9] which can be tested via experiments or observations. Because Eqs.(1,2) are covariant and have substantial physical meanings, the author believes that the Lorentz and Levi-Civita's conservation laws might be better than Einstein's conservation laws and tries to use the Lorentz and Levi-Civita's conservation laws to study black holes. Of course, whether these views are correct should be tested by future experiments and observations.

Evidently Eqs.(1,2) can be rewritten as

$$\frac{\partial}{\partial x^{\lambda}}\left(\sqrt{-g}\,T^{\lambda\sigma}_{(M)} + \sqrt{-g}\,T^{\lambda\sigma}_{(G)}\right) = 0 \qquad (1')$$

$$\sqrt{-g}\,T^{\lambda\sigma}_{(M)} + \sqrt{-g}\,T^{\lambda\sigma}_{(G)} = 0 \qquad (2')$$

The Einstein's field equations [10]

$$R^{\mu\nu} - \frac{1}{2} g^{\mu\nu} R = -8\pi G\,T^{\mu\nu}_{(M)} \qquad (5)$$

is one important theoretical foundation to study black holes. From Eqs. (1',2',5) we obtain

$$T^{\mu\nu}_{(G)} + T^{\mu\nu}_{(M)} = 0 \qquad (6)$$

$$\frac{\partial}{\partial x^{\mu}}(T^{\mu\nu}_{(G)} + T^{\mu\nu}_{(M)}) = 0 \qquad (7)$$

where $T^{\mu\nu}_{(G)} = \frac{1}{8\pi G}(R^{\mu\nu} - \frac{1}{2}g^{\mu\nu}R)$. Let $\mu = \nu = 0$ in Eqs. (6,7) then we have

$$\rho_G + \rho_M = 0 \qquad (8)$$

$$\frac{\partial}{\partial t}(\rho_G + \rho_M) = 0 \qquad (9)$$

where $\rho_M$ is the energy density of the matter field, $\rho_G$ is the energy density of the gravitational field. If we assume that in all circumstances $\rho_M \geq 0$, then there is $\rho_G \leq 0$ always.

From Eq. (8) we get $\triangle \rho_M = -\triangle \rho_G$ immediately, this relation means that for an isolated gravitational system if the energy density of the matter field increases, then the energy density of the gravitational field should decrease, *i.e.* the energy density of the gravitational field might transform into the energy density of the matter field. Inversely if the energy density of the matter field decreases, then the energy density of the gravitational field should increase, *i.e.* the energy density of the matter field might transform into the energy density of the gravitational field. Since $\Delta \rho_M + \Delta \rho_G = 0$, hence when $\Delta \rho_M > 0$ then $\Delta \rho_G < 0$; for these cases both the positive value energy density of $\Delta \rho_M$ and the negative value energy density of $\Delta \rho_G$ might appear to be originating from zero energy. It is possible for these changes to happen, an extreme example is given in Ref. [3] which has shown that the energies of whole matter field in the cosmos might originate from the gravitational field. On the other hand, when $\rho_M > 0$, $\Delta \rho_M < 0$ then $\rho_G < 0$, $\Delta \rho_G > 0$; for these cases some of the positive value energy density of $\rho_M$ and some of the negative value energy density of $\rho_G$ might mutually annihilate each other to zero energy. We shall show in section 5 that these changes could also happen.

## 3. The entropy of a closed system

The second law of thermodynamics tells us that, for every physics process in a closed system the entropy change $\Delta S$ of this system must satisfy the relation [11,12]:

$$\Delta S \geq 0 \tag{10}$$

The entropy $S$ obeys Boltzmann's relation [11,12]

$$S = k \ln N \tag{11}$$

where N is the number of microscopic states of this closed system. If the whole closed system can be divided into a number of subsystems, then [12]

$$N = \prod_a \Delta N_a \tag{12}$$

$\Delta N_a$ is the number of microscopic states of the subsystem denoted by $a$. In the quasi-classical case,

$$\Delta N_a = \Delta p_a \Delta q_a / h^s \tag{13}$$

where $\Delta q_a$ represents the interval of coordinates, $\Delta p_a$ represents the interval of momentums, $h$ is Plank constant, $s$ is the number of degrees of freedom of the subsystem denoted by $a$ [12].

For a star, the material of the star and its gravitational field constitute a closed system. The calculations of microscopic states for a gravitational system need a quantum theory of a gravitational field. As a complete and consistent quantum theory of a gravitational field has not yet been constructed, we can only make some rough estimates. In section 5 we shall use Eqs.(10-13) to discuss qualitatively the formation process of a black hole.

## 4. The event horizon of a black hole

In studying the formation process of a black hole, we might use the Schwarzchild metric [10,13,14]:

$$ds^2 = -(1-\frac{2M}{r})dt^2 + \frac{1}{1-\frac{2M}{r}}dr^2 + r^2(d\theta^2 + \sin^2\theta\, d\phi^2) \qquad (14)$$

and the Kerr metric [13,14]:

$$ds^2 = -(1-\frac{2Mr}{\rho^2})dt^2 + \frac{\rho^2}{\Delta}dr^2 + \rho^2 d\theta^2 + [(r^2+a^2)\sin^2\theta + \frac{2Mr\,a^2\sin^4\theta}{\rho^2}]d\phi^2 - \frac{4Mr\,a\,\sin^2\theta}{\rho^2}dt\,d\phi \qquad (15)$$

where $\rho^2 \equiv r^2 + a^2\cos^2\theta$, $\Delta \equiv r^2 - 2Mr + a^2$, $a = \frac{J}{M}$, M and J are the mass and the angular momentum of the star respectively. In Eqs. (14, 15) the natural unit system is used hence c=1 G=1.

The radius of event horizon of a Schwarzchild black hole is

$$r_g = 2M \qquad (16)$$

which can be derived from Eq.(14) [10,13,14]; the radius of event horizon of a Kerr black hole is

$$r_{\pm g} = M \pm \sqrt{M^2 - a^2} \qquad (17)$$

which can be derived from Eq.(15) [13,14].

## 5. The changes in the formation process of a black hole

In the formation process of a black hole from a star, the star collapses upon itself due to its gravitational contraction. It contracts from a radius which is greater than the radius of event horizon denoted by Eq. (16) or Eq. (17). When the star has shrunk into its event horizon, it becomes a black hole. Everything can pass in through the horizon, but it cannot get back out.

For a Schwarzchild black hole, once some substances (including some parts of the star itself) have crossed the horizon, they are doomed to move inexorably closer and closer to the 'singularity' at the center of the black hole. For a Kerr black hole, once some substances (including some parts of the star itself) have crossed the outer horizon, they are doomed to move inexorably closer to and cross the inner horizon of the black hole.

Suppose a contracting star has sufficient mass M**,** we first discuss that the metric of the star is Schwarzchild metric. Eq. (16) tells us that when the radius $r$ of the star is reduced to $r = (r_g)_M = 2M$, the event horizon should appear. If in the meantime just at the radius of the star being reduced to $r = (r_g)_M = 2M$, the mass M of the star decreases to $M - \Delta M$. The radius of event horizon corresponding to $M - \Delta M$ is $(r_g)_{M-\Delta M} = 2(M - \Delta M)$. Evidently

$$(r_g)_M > (r_g)_{M-\Delta M} \tag{18}$$

Hence when the radius of the star is reduced to $r = (r_g)_M = 2M$ but its mass is $M - \Delta M$, the event horizon would not appear.

The Einstein's conservation laws maintain that the mass of a star cannot decrease spontaneously, but the Lorentz and Levi-Civita's conservation laws maintain that the mass of a star could decrease spontaneously under some conditions. The possibility of spontaneous decrease of mass is due to the relation $\Delta \rho_M + \Delta \rho_G = 0$, which means it is probable that some positive value energy density of $\rho_M$ and some negative value energy density of $\rho_G$ might be annihilated mutually to zero energy. Since according to the theory of quantum mechanics, black holes might not exist; hence we could suppose that when the radius r of the star approach to $(r_g)_M = 2M$ from $r > (r_g)_M = 2M$, the mass M of the star would decrease continuously such that the Eq. (18) is always satisfied when M takes arbitrary value. Therefore the event horizon would not appear and the black hole would not be formed. This supposition might be interpreted as "the Lorentz and Lavi-Civita conservation laws prohibit the existence of black holes".

If the metric of the star is Kerr metric, let $M + \sqrt{M^2 - a^2}$ instead of 2M and

$$(r_g)_{M+\sqrt{M^2-a^2}} > (r_g)_{M-\Delta M + \sqrt{(M-\Delta M)^2-a^2}} \tag{18'}$$

instead of Eq.(18), we can reach the same conclusion.

When a part of the star mass is annihilated, the number of subsystems for this star will decrease. From Eqs.(11,12) the entropy of this star will also decrease. Why can the annihilating process take place? It must be pointed out that whenever a part of star mass is annihilated, there are gravitational waves emitted at the same time. In this case the closed system being considered must include the gravitational waves, and then the entropy of the whole closed system would be possible to increase in the process. The vast majority of the stars have angular momentum and their metric are Kerr metric; they will emit gravitational waves whenever a part of star mass is annihilated and the radius of the star is decreased, hence no further discussion is necessary. However if the metric of the star is Schwarzchild metric, then according to Birkhoff theorem [10] it is not possible to emit gravitational waves for a contracting star. But it must be indicated that there are two preconditions for Birkhoff theorem: 1, the mass of a star must keep constant; 2, the exterior of a star is vacuum always, *i.e.* any matter field does not exist there, hence we may use the Einstein field equation in vacuum: $R^{\mu\nu} - \frac{1}{2}g^{\mu\nu}R = 0$. If the Einstein's conservation laws are correct, these two preconditions are tenable; but if the Lorentz and Levi-Civita's conservation laws are correct, these two preconditions are untenable, and the Birkhoff theorem does not hold water. Since when the mass of a star changes, the Schwarzchild metric Eq.(14) must also change, *i.e.* the gravitational waves are radiated. We must also indicate that, owing to Eq.(2), the exterior of a star can't be vacuum always.

Finally we must emphasize that the condition for the annihilation of a part of star mass is "when the event horizon would appear". If there is no event horizon, the annihilation does not take place spontaneously, as it had been explained in Ref.[7].

## 6. Further discussions

By using the Lorentz and Levi-Civita's conservation laws it is shown in this paper that the nonexistence of black holes could be explained by the theory of classical gravity.

Hawking-Penrose singularity theorem [15] has been extensively used to explain the existence of blackholes. It is noted that the existence of a closed trapped surface is one precondition[15] of Hawking-Penrose singularity theorem. The event horizon of black hole is a closed trapped surface. The nonexistence of event horizon would mean that the closed trapped surface might not exist, therefore the singularity theorem would not establish.

The definitive answer to the existence of black hole should be decided by experiments and observations. The theoretical studies of black holes have already reached a very extensive stage, but black hole hasn't been observed yet and there appear to be some un-resolvable difficulties in its theories, especially there exists contradiction between event horizon and quantum mechanics. Therefore, it is worth to explore the thinking that the hypothesis of non-existence of black holes would be a logical way out for fundamentally resolving the contradiction and difficulties. Investigations of the theories of the non-existence of black hole have just been started and are still in an early stage. Further and more careful investigations of these and other theories, as well as thorough discussions with researchers of opposition standpoint, would promote more understanding of the black hole theories and the design of methods to test its existence or non-existence.

If the black holes do not exist, the explanations of some astrophysical phenomena by using black hole theories must be reconsidered. Regardless of whether or not black hole exists, the existing theories of black hole and thermodynamics of black hole provide interesting mathematical models.